# The influence of sponsors on organizational structure of free software communities
## Invited Paper

Daniel Esashika
University of São Paulo
Faculty of Economy, Management and Accountability
+55 (11) 3091-5960
daniel.esashika@usp.br

Carlos Denner dos Santos Junior
University of Brasília
Faculty of Economy, Management and Accountability
+55 (61) 3107-0759
carlosdenner@unb.br

**ABSTRACT**

Initially, free software communities are characterized by self-management, however, they were also influenced by public and private organizations that identified potential gains in the use of the geographically distributed production model. In this context, this research aims to answer the following questions: Do sponsors influence the organizational structures of free software communities by promoting differences between sponsored and non-sponsored communities? What strategies are adopted by the sponsor to influence the organizational structure of free software communities? Two constructs are central to the study: organizational structure and sponsorship. For this research, we adopted case study methodology and three free software communities were studied. In the analysis of the results it was evidenced that sponsors influence decision making, definition of community key roles, and a formalization of norms. In turn, non-sponsored communities were characterized by the centralization and informality of the norms. We conclude that differences were identified in the organizational structure of sponsored and non-sponsored free software communities, and this differentiation was influenced by sponsors. In addition, it was possible to describe strategies and mechanisms used by sponsors to influence the community organizational structure.

## 1. INTRODUCTION

Virtual communities as types of organization are influenced by groups which they maintain relationships (Pfeffer & Salancik, 1978). The relevance of each of these groups is related to their capability to influence the organization's objectives (Mitchell *et al.*, 1997). The sponsors are key players in understanding for virtual communities, because they have essential resources to maintain the organization (Hillman *et al.*, 2009).

Examples of virtual communities are communities of practice, consumer communities, activist communities, and software communities (Kozinets, 1999; Brown & Deguid, 2001; O'Mahony & Ferraro, 2007; Lee, 2008). Free software communities stand out in this group, especially because their level of sophistication in governance and collaborative work (O'Mahony, 2001; West & O'Mahony, 2005; O'Mahony & Ferraro, 2007; O'Mahony, 2007). A free software can be executed, adapted, redistributed, and refined by users (Hill *et al.*, 2007). Initially, free software communities were characterized by self-management. Currently, they were also influenced by public and private organizations that identified potential gains in using the collaborative and geographically distributed production model (Fitzgerald, 2006; O'Mahony, 2007). (Riehle & Berschneider, 2012).

In this sense, the studies began to investigate the influence of sponsoring organizations on the organizational structure of free software communities (O'Mahony, 2007). There is evidence that sponsors of software communities are responsible for strategic decisions (Stuemer, 2009), division of tasks (van Wendel de Joode, 2004, Wijnen-Meijer & Batenburg, 2007) and the formalization of norms (Lakhani & Wolf, 2005; Blekh, 2015), contrasting with the philosophy of independence and collaborative management that characterized the early software communities (Raymond, 1998).

In this context, our research aims to answer the following questions: Do sponsors influence the organizational structures of free software communities, promoting differences between sponsored and non-sponsored communities? What strategies are adopted by the sponsor to influence the organizational structure of free software communities? This study presented strategies and mechanisms used by sponsors in interventions in the free software communities, as well showed organizational differences between sponsored and non-sponsored communities. The clarification of this relationship allows to understanding the phenomenon of expansion of sponsored free software communities (West & O'Mahony, 2005, Stuermer *et al.*, 2009, Androutsellis-Theotokis *et al.*, Nielsen *et al.*, 2003).

## 2. ORGANIZATIONAL STRUCTURE

There are many definition to organizational structure, but in general is defined as a form of coordination and control through administrative mechanisms, in order to integrate organizational units that perform different activities (Lawrence & Lorsch, 1967; Jackson & Morgan, 1982). The coordination and control of the organizational activities are operationalized by hierarchical levels, mechanisms of integration between the areas, distribution of roles and responsibilities, and the formalization of norms. (Thompson, 1967; Walton, 1986; Sablynski, 2012). From the definitions of organizational structure established above, we characterized structure based on three characteristics: centralization of decision-making, division of labor and specialization, and formalization of the organization's procedures (Hall, 1967; Zheng, *et al.*, 2010; Mafini, 2014; Ho *et al.*; 2014; Jorge & Carvalho, 2014; Foss *et al.* 2014; Lin, 2012; Worley & Doyle, 2015; Gibson *et al.*, 2015).

Centralization is related to the concentration of authority in decision making in one person, department, or hierarchical level (Galbraith, 1974; Ferell & Skinner, 1988; Jensen & Meckling, 1992; Schminke *et al.*, 2002). For Simon (1977), identifying a management of decision-making is essential to understand the organization. An organization with a high degree of decentralization, authority is delegated to all levels and employees are free to perform their duties in the way they consider most productive (Andrews *et al.*, 2008). Otherwise, a high degree of centralization or top-down control, when there is concentration of authority in a higher instance in the organization (Andrews *et al.*, 2008). In literature, opposite results have been reported. Some authors have found positive results with centralization (Ruekert *et al.*, 1985) and others have obtained better results with decentralization (Burns & Stalker, 1961; Dewar & Werbel, 1979; Chen & Huang, 2007). Centralization facilitates performance control, making it more predictable (Germain *et al.*, 2008). On the other hand, it has several disadvantages cited in the literature, such as difficulty in maintaining communication, commitment and employee involvement (Chen & Huang, 2007). On the other hand, decentralization favors the exploitation of market opportunities, since work processes are more flexible and adaptable, as well as favor the communication and perception of employees of a good work environment (Burns & Stalker, 1961; Dewar & Werbel, 1979; Armandi & Mills, 1982; Stuart & Podolny, 1996; Schminke *et al.*, 2002; Smith & Tushman, 2005; Tran & Tian, 2013).

The division of labor is a distribution of activities in an organization, being influenced by the differentiation of the tasks and the specialization of the areas (Kahn *et al.*, 1964), or a consequence of the organization's growth (Blau, 1970). There is controversy over the impact of organizational size on the differentiation of structures, and some studies have pointed to this relationship as irrelevant (Hall *et al.*, 1967). The literature on organizational theories highlights this trade-off between departmentalization and coordination costs. In other words, more specialization and differentiation implies higher costs of coordination (Blau, 1970; Weber, 2004). Therefore, one of the consequences of the differentiation of structures is the increase of the administrative component of organizations, areas specialized in managing the interdependence of the work of other sectors (Blau, 1970). In addition, the division of labor is a rational positioning of the organization against a heterogeneous task environment, seeking in this context to identify homogeneous segments and establish units in the organizational structure to deal with each specific segment (Thompson, 1967).

Finally, formalization is the evaluation of the use of rules and procedures to guide users' behaviors and the decision-making process in the organization (Zheng *et al.*, 2010). In addition, formalization indicates how much of the principles, policies, procedures, and rules for managing firm processes are formally registered (Lee & Choi, 2003; Pertusa-Ortega *et al.*, 2010). When an organization has a high degree of formalization, an execution of business processes is well described and written (Willem & Buelens, 2009). There are divergences in the literature about the effects of formalization. For some authors, organizations with a high degree of formalization constrain the spontaneity and flexibility necessary to improve communication and internal interaction (Nonaka & Takeuchi, 1995; Chen & Huang, 2007), while others demonstrate that formalized organizational structures contribute to the organizational effectiveness (Wang, 2003). For instance, formalization allows members of the organization to understand the productive flow within the company, thus facilitating member cooperation, collaboration, trust (Schminke *et al.*, 2002; Jansen *et al.*, 2006), decision making and communications (Ferrell & Skinner, 1988; Nahm *et al.*, 2003).

## 3. SPONSORSHIP

The proposals under discussion are related to the sponsor, a specific player who invests resources in the development of the organization. Madill & O'Reilly (2010) define sponsorship with two characteristics: an association and existence of mutual benefits in the exchange of resources between sponsors and sponsored. Sponsorship can take place in a variety of ways, such as through government intervention, development agencies, universities, profit and non-profit organizations (Flynn, 1988), individuals (Karpoff *et al.*, 1996).

Sponsors are relevant stakeholders because they have the capacity to influence the firm's objectives or to control essential resources (Mitchell *et al.*, 1997). Studies on the topic have highlighted the influence of sponsors in organizations' governance decisions, particularly institutional sponsors, who have greater decision-making power over individual sponsors (Gillian *et al.*, 2000). In addition, we highlight the studies that demonstrate the influence of the sponsors in decisions related to the research and development of firms (David *et al.*, 2001, Lee & O'Neill, 2003) and in decisions about policies and norms of the organizations (Galant, 1990; Karpoff *et al.*, 1996). This influence is usually effected by acting with the managers of the sponsored organization (Karpoff, 2001).

In this sense, production studies in sponsored free software communities point out that community sponsors tend to get involved in relevant decisions for software development (Stuemer, 2009). On the other hand, in non-sponsored communities, decisions tend to be more collaborative (Raymond, 1998; O'Mahony, 2007), even though this aspect is not consensual among authors. For some authors, centralization of decisions is also common in non-sponsored communities, with the roles of the project manager and core group members being highlighted in the decision-making process (Ye & Kishida, 2002; Crowston & Houston, 2006; Dafermos, 2012).

Other relevant aspect also has been considered is the formalization of the division of labor in free software communities, because it is important to sponsors to establish clear rules of division of labor for the investments rationalization. Free software communities enable the parallel development of multiple versions of the software and the modularization of software into specific functionalities (van Wendel de Joode, 2004). In this sense, a clearer and more formal definition of the possibilities of action and division of labor favors the saving of resources to the sponsoring firms, since they allow the targeted allocation to versions and modules that favor their interests (Krishnamurthy, 2003; Henkel, 2006; Wijnen-Meijer & Batenburg, 2007).

The sponsors have been working with the managers of the sponsored organizations in the elaboration of policies and norms, in order to guarantee their investments (Galant, 1990, Karpoff *et al.*, 1996, Karpoff 2001, Lakhani & Wolf, 2005). In this sense, sponsored free software communities have similar characteristics to other sponsored organizations, with sponsors intervening to establish governance rules (Androutsellis-Theotokis *et al.*, 2010; Blekh, 2015). In turn, non-sponsored communities tend to be characterized by the bazaar model, described as management based on the flexibility and autonomy of the collaborators (Raymond, 1999).

## 4. METHODOLOGY

In order to investigate the influence of the sponsors in the organizational structures of the free software communities, a descriptive research was carried out using secondary data obtained from repositories of free software projects, which served as complement to semi-structured interviews with members of the software communities and employees of the sponsoring companies.

Cases

The cases have been chosen to illustrate a non-sponsored community (GoboLinux) and sponsored communities, one by cooperative (Noosfero/Colivre) and by government (Portal Modelo/Interlegis). Which one are more than a decade old, demonstrating the stability of the community, thus allowing the identification of the characteristics of its development and organization models (Androutsellis-Theotokis, 2010). Finally, the communities were identified as active, demonstrating that members continue to collaborate in the functioning of the community. In addition, selected communities were started in Brazil, one of the countries in which the free software movement has gained more relevance due to the number of developers involved and the importance of the events carried out (Evangelista, 2014). In addition, the Brazilian government has a policy for the development and stimulation of the use of free software, with emphasis on the Executive Committee on Electronic Government (Decree No. 29, 2003) and the Public Software Portal, which has more than 60 solutions In Brazil and other Latin American countries (Brazil, 2016).

Categories

The organizational structures of the software communities analyzed in this case study were classified according to analytical categories from the review of literature. Table 1 summarizes the analytical categories used in this study

Documentary research

The data were collected electronically and three secondary sources of data were used: the documentation and reports on the project, messages sent to the software forum and messages sent to the public mailing list. Reading these messages allowed the familiarization with the types, quantities, contents and specific contributions of each developer, helping to understand the dynamics of the functioning of the community studied. This stage of the research had the objective to analyze aspects related to the influence of the sponsors in the structure of the software communities in formal terms manifested in documentation and reports.

Interviews

The interviews were conducted with a semi-structured guide, through direct contact or videoconference. Each interview had an approximate duration of 40 minutes and aspects such as the nature of the work performed, the sponsor's priority interests, the project's history, the conflicts of interest between those involved in the project and the nature of the interactions between sponsoring organizations and volunteer developers.

Participants of the interviews were selected from the core group participants and members of the free software communities whether volunteers, employees or members of the sponsoring organizations. The interviews with core group members were particularly important because of their influence on the organization and structure of the community. The number of interviews was determined by the theoretical saturation, that is, the point at which new interviews do not aggregate different information and categories (Bauer & Gaskell, 2000).

Procedures for data analysis

The documents, obtained directly from the documentary research and the interview transcripts, were organized by case. Each interviewee's speech was analyzed and classified according to the potential to elucidate the points referring to the three analytical categories of structures described in this research. The classified texts were separated and compared for the selection of those that best explain the evaluated aspects (Bardin, 1979; Corbin & Strauss, 1990).

## 5. RESULTS

On the centralization of decision-making, we found different results for sponsored and non-sponsored communities. In sponsored communities, there is a clear leadership role played by community members who are related to sponsors. This characteristic corroborates the observations of Shah (2006), who found that the work teams kept by the sponsors tend to define the directions that the community should take.

> Colivre is fundamental to the Noosfero community because it is a major funder of the project. The main developers and current engineers are formed directly or indirectly by Colivre (Member of Noosfero's Core Group).

However, community members recognize the leaders' intention to keep the decision-making process as decentralized and consultative as possible, according to the values of free software communities, such as democratic and shared management (Raymond, 1999). This characteristic is identified in the interviewees 'speech, which recognizes in the representatives of the sponsoring entity the interest in receiving and analyzing the suggestions that come from the community, strengthening the members' motivation and sense of belonging (Hinds & Lee, 2008).

**Table 1. Analytical Categories and Description. Source: authors.**

| Categories | Description | Example |
|---|---|---|
| Centralization | Centralized: decisions are taken by the top managers (Moch & Morse, 1977; Ashmos *et al.*, 1998; Moynihan & Pandey, 2005; Lin, 2012). | Prevalence of decisions made by the sponsor, project leader, or community core group. |
| | Consultive: the decisions are taken by the top managers with validation and contributions from the community. | Decisions made by the sponsor, project leader or community core group, heard from the community. |
| | Decentralized: decisions are taken predominantly by the community (Moch & Morse, 1977; Ashmos *et al.*, 1998; Moynihan & Pandey, 2005; Lin, 2012). | Prevalence of decisions taken by community developers |
| Division of Labor | Departmentalized: the organization presents differentiation, observed by the quantity and precision of divisions or departments (Hall *et al.*,1967; Zeffane, 1992; Yang, 2008; Mafini, 2014; Gibson *et al.*, 2015). | Existence of committees, councils, specific security teams, release of versions, communication, documentation, quality management or maintenance. |
| | Role segmentation: the organization presents a rudimentary form of differentiation, in which some roles are already well defined (Ye & Kishida, 2002). | Presence of project leader, core group, active developers, peripheral developers, bug reporters, bug fixes, readers, passive users (Ye & Kishida, 2002). |
| | Undifferentiated: the organization does not show differentiation, observed by the quantity and precision of divisions or departments (Hall *et al.*,1967; Zeffane, 1992; Yang, 2008; Mafini, 2014; Gibson *et al.*, 2015). | Voluntary non-systematic or standardized development, without organizational relationship. |
| Formalization | Formal-analytic: prevalence of rules, procedures, policies and standardizations registered in documents (Yang, 2008; Ho *et al.*, 2014). | Development policies, minutes of decisions, social contracts, description of members' roles, organization chart, codes of conduct, manuals. |
| | Formal-synthetic: prevalence of rules, procedures, policies, and standards that are not documented. | Minimum set of registered procedures and policies, such as the Development Policy and rules for the establishment of new members in the core group. |
| | Informal: lack of registration of rules, procedures, policies or standards (Yang, 2008; Ho *et al.*, 2014). | Content of unregistered informal conversations, associated with the memory of the members who participated in the events and decisions. |

I would not say they are centralizers, they are responsible. Communities must have diverse voices, the role of leadership is to encourage discussions and make decisions that are aligned with the proposed long-term vision. (Member of Portal Modelo's community).

In turn, in the non-sponsored community, there were signs of centralization in the decision-making process. One possible explanation for this fact is the small size of the developer group, with centralization in the core group being a natural mechanism for maintaining mutual trust. In the case under study, the group that maintained the community was previously known in a face-to-face environment, transporting the established relationships in an offline environment to the virtual community.

Regarding the division of labor, sponsored and non-sponsored communities presented an initial stage of departmentalization, characterized by the segmentation of participants in roles. However, it is important to highlight the importance of understanding who holds the positions with decision-making power in each case. In sponsored communities, it should be noted that the role of community coordinator and developers with access to the core code was filled by developers contracted by the sponsor or institutional partners of the sponsor. In the case of the non-sponsored community the role of coordinator was occupied by the creator of the application and the positions of developers with access to the code were granted to users with more contributions to the software code. Therefore, it can be inferred that one of the strategies of intervention of the sponsor in the structure of the community is the distribution of the decision-making roles of the community among its contracted employees, providing a greater control over the development carried out within the community.

Regarding the formalization of the norms, the projects maintained by sponsors differ from the ones not maintained due to the formalization of the norms. Although other studies point to informality as a feature of free software communities (Henkel 2006, Sadowski *et al.*, 2008), in this study, sponsored communities implemented formal rules and policies.

> Only the new functionalities that are in the roadmap of the product will be implemented, according to milestone registration. All new functionality proposed by the community should be preceded by discussion in the respective -dev list of the product and should have its insertion previously agreed upon. The insertion can happen either through the opening of a new ticket or the re-assignment of an existing ticket (Portal Modelo/ Interlegis - Development Policy, 2015).

In opposite, non-sponsored community was characterized as informal. This is because community members find it unnecessary to formalize standards for a small group of developers. From the interviewees' speech, the emergence of norms in free software communities is a consequence of the emergence of managerial demands, corroborating the position in the literature that governance formalization is not an imposition, but rather a mechanism that emerges from the shared perception of a need (West & O'Mahony, 2005). Particularly, this was clear in the case of a sponsored community that established formal norms because of new organizational actors into the community.

> This document was a consequence of the approach of new agents, besides Colivre, in the project community. It was a work done by Colivre with the intention of documenting and making feasible the process of entry of these new agents in the community, besides defining some rules of community functioning, which were previously dispersed in the group's own culture, more explicitly (Member of Noosfero's Core Group).

Table 2 summarizes the results obtained for each analytical category in the cases studied in this study.

**Table 2. Comparative table of case results. Source: authors.**

| Categories | Portal Modelo / Interlegis | Noosfero / Colivre | GoboLinux |
|---|---|---|---|
| Type of Sponsor | Sponsored by public organization | Sponsored by public cooperative | Non-sponsored |
| Centralization | Consultive | Consultive | Centralized |
| Division of labor | Role segmentation | Role segmentation | Role segmentation |
| Formalization | Formal-Synthetic | Formal-Analytic | Informal |

## 6. CONCLUSIONS

From the researched literature, three analytical categories were identified to describe the organizational structure of free software communities: centralization, division of labor and formalization. Based on these categories, three case studies involving sponsored and non-sponsored free software communities were carried out to analyze the influence of sponsors in the organizational structures of the communities, identifying structural differences between the communities and the strategies of the sponsors.

In the sponsored communities it was evidenced that consultative character gives legitimacy to sponsor's role in the decisions that are made within the community. Regards to the division of labor, sponsors play a role in defining the key roles of the community, which have the power of decision making, and it is essential for the community's organization and achievement of sponsors' objectives. Finally, the sponsors also influence the formalization of community norms as a mechanism to clarify roles and work flows, legitimizing their interests towards volunteer developers and facilitating the negotiation for new organizational participants.

In contrast to the sponsored communities, this study presented the case of a non-sponsored free software community which is characterized by the centralization of decisions in the group of developers. Also, the community was characterized by segmentation in roles, and informalization of community norms and guidelines. Centralization of decisions and the informalization of norms are justified by the small size of the non-sponsored community support group. In addition, although the study communities have the same approximate time of existence, the difference in organization and resources of the sponsored free-software communities, which are distinguished by the greater capacity of management and development, is remarkable.

For theory, the study presents differential elements between sponsored and non-sponsored free software communities, which are influenced by the sponsor's performance in their interaction with the community. For practical purposes, the study presents elements that can be used by organizations that are interested in investing in software developed by the free software community, which can help these organizations and managers achieve greater legitimacy and acceptance in free software communities.

As future research, we recommend three questions that have emerged throughout the research. First, we suggest investigating the relationship between sponsorship and the sustainability of open source projects. During the research, the role of the institutional sponsor for maintenance of open projects was becoming increasingly evident. Second, the sharing of structures between sponsored organization and sponsor. In our observations, these initiatives were consolidated through foundations and associating with other institutions to ensure their sustainability. Third, another aspect that should be addressed in a research is a typology of the types of sponsors of free software communities.